\pdfoutput=1

\documentclass[11pt]{article}

\usepackage[preprint]{acl}

\usepackage{times}
\usepackage{latexsym}

\usepackage[T1]{fontenc}

\usepackage[utf8]{inputenc}

\usepackage{microtype}

\usepackage{inconsolata}

\usepackage{algorithm}
\usepackage{algpseudocode}
\usepackage{graphicx}
\usepackage{subfigure}
\usepackage{stfloats}

\usepackage{amsmath}
\usepackage{bbding}

\usepackage[marginal]{footmisc}

\newcommand{\secref}[1]{Sec. \ref{#1}}
\newcommand{\figref}[1]{Figure \ref{#1}}

\newcommand{\tabref}[1]{Table \ref{#1}}

\title{Phonetic and Lexical Discovery of a Canine Language using HuBERT}

\author{Xingyuan Li\textsuperscript{1}\footnote[1], Sinong Wang\textsuperscript{2}\footnote[1], Zeyu Xie\textsuperscript{3}, Mengyue Wu\textsuperscript{4}\footnote[2], Kenny Q. Zhu\textsuperscript{5}\footnote[2] \\ \\ 
  \textsuperscript{1,3,4}Shanghai Jiao Tong University, Shanghai, China \\
  \textsuperscript{2,5}University of Texas at Arlington, Arlington, Texas, USA \\
  \texttt{\{\textsuperscript{1}xingyuan, \textsuperscript{3}zeyu\_xie, \textsuperscript{4}mengyuewu\}@sjtu.edu.cn}\\
  \texttt{\textsuperscript{2}sxw7663@mavs.uta.edu}\\
  \texttt{\textsuperscript{5}kenny.zhu@uta.edu}\\}

\begin{document}

\maketitle
\footnote[1]{\textsuperscript{*}These authors made equal contribution.}
\footnote[2]{\textsuperscript{\dag}Corresponding authors.}
\begin{abstract}

This paper delves into the pioneering exploration of potential communication patterns within dog vocalizations and transcends traditional linguistic analysis barriers, which heavily relies on human priori knowledge on limited datasets to find sound units in dog vocalization. 
We present a self-supervised approach with HuBERT, enabling the accurate classification of phoneme labels and the identification of vocal patterns that suggest a rudimentary vocabulary within dog vocalizations. Our findings indicate a significant acoustic consistency in these identified canine vocabulary, covering the entirety of observed dog vocalization sequences. 
We further develop a web-based dog vocalization labeling system. This system can highlight phoneme n-grams, present in the vocabulary, in the dog audio uploaded by users.

\end{abstract}

\section{Introduction}

The concept of animal language hinges on the intricate ways through which non-human species communicate, revealing a spectrum of vocalizations, gestures, and behavioral cues that parallel human efforts to convey information and emotions. The study of animal communication has captivated researchers across numerous disciplines,
from biology to linguistics~\citep{rutz2023using,paladini2020bark,robbins2000vocal}. 
While it is widely accepted that most animals lack a language system comparable to human language,
recent advancements in natural language processing have opened up new avenues for investigating the patterns and structures embedded within animal vocalizations~\citep{huang2023transcribing,wang2023towards}. The study of animal language, therefore, not only broadens our understanding of the animal kingdom but also deepens our insights into the evolution and functions of communication systems across species.

\begin{figure}[th]
    \centering
    \scalebox{0.35}{\includegraphics{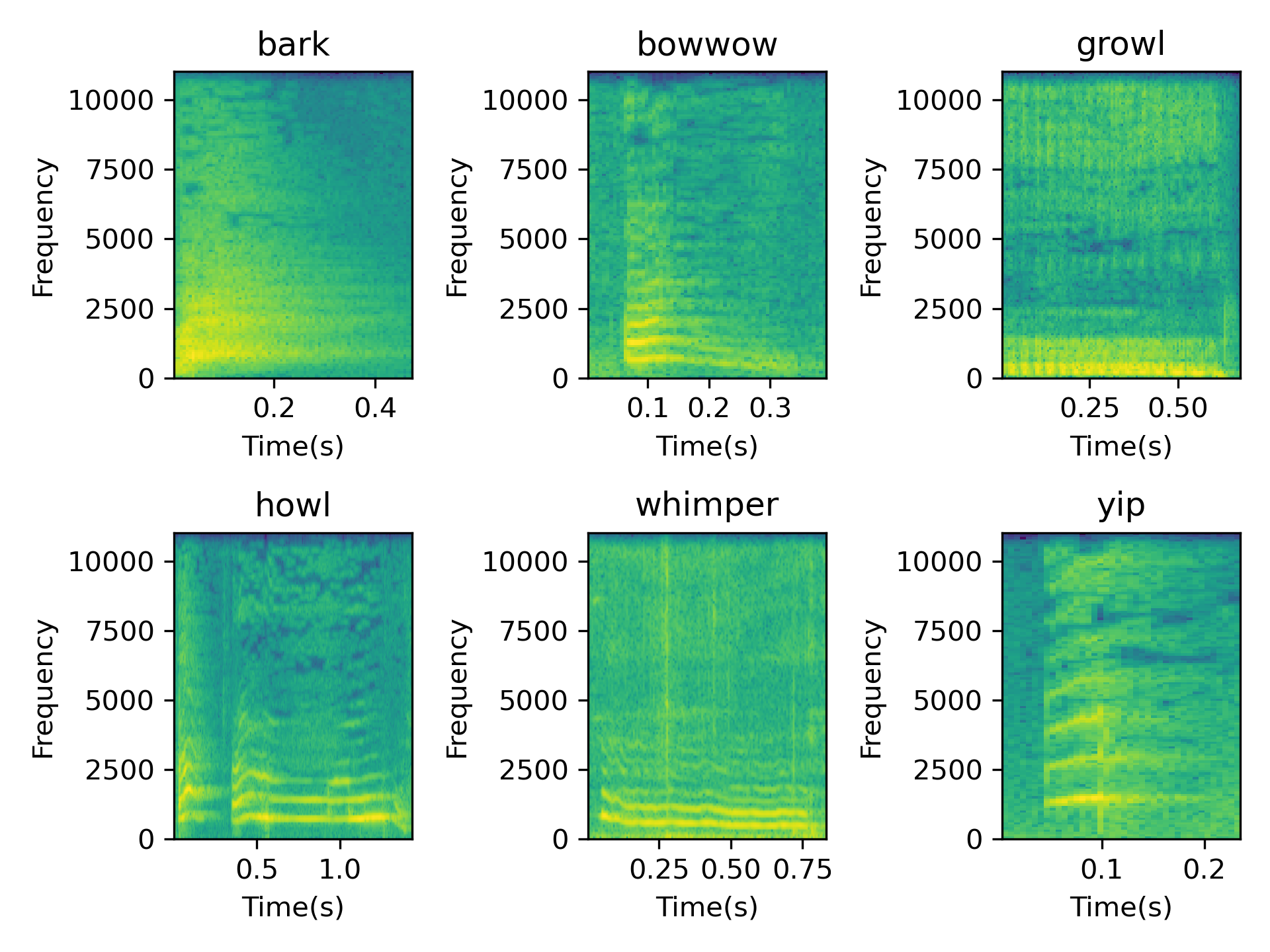}}
    \caption{Six different dog barking sounds from AudioSet~\citep{gemmeke2017audio}}
    \label{fig:six_word}
\end{figure}

Despite observable acoustic variations in animal vocalizations that hint at potential patterns of communication, asserting that animals, such as dogs, possess a language is fraught with limitations. The absence of identifiable phonemes and structured syntax in their vocalizations challenges traditional linguistic frameworks~\citep{holdcroft1991saussure}. While dogs exhibit a range of sounds that vary in pitch, duration, and intensity, as illustrated in Figure \ref{fig:six_word} (sound clips extracted from Audioset~\citep{gemmeke2017audio}), six distinct dog barking sounds are identified. However,  these variations alone do not constitute a language. The lack of a structured, consistent phonetic system and the inability to form complex ideas through their sounds significantly limit the comparison of their vocalizations to human language.




Nevertheless, undertaking the challenge of exploring the concept of animal language is a daunting task. Different from well-researched human language, the frequency range, and phonetic variations remain under-discovered, rendering the classification approach based on sound amplitude inadequate for discerning the fundamental phonemes of dog vocalizations. The difficulty lies not only in interpreting the acoustic variations but also in identifying meaningful patterns within the vast array of animal sounds. This complexity is compounded by the need to distinguish between mere noise and significant vocalizations that could indicate some form of structured communication. The endeavor requires innovative methodologies and a departure from traditional linguistic analysis. 
Our approach to navigating these challenges involves leveraging advanced signal processing techniques and self-supervised learning models. By focusing on the acoustic features of dog vocalizations, we aim to uncover underlying patterns that could suggest a rudimentary form of communication. This involves a meticulous process of audio clean-up, sentence extraction, phoneme recognition and combination, and the establishment of word discovery across vocalizations from different individual dogs. Given the lack of prior knowledge of dog vocalizations, we apply a self-supervised approach HuBERT~\citep{hsu2021hubert} for pretraining and phoneme identification.
HuBERT can effectively reference the context information of the audio and generate vector representations, which enables it to generate stable results when faced with vocalizations like language that may have slight variations in context. 

Utilizing the precise phoneme labels acquired, we explored the feasibility of creating a vocabulary from the available dog sound dataset, characterized by complete and commonly occurring phoneme sequences vocalized by numerous dogs, which could be identified as a ``word''.
We developed popularity score to quantify the likelihood of a phoneme ngram constituting a ``word'', assessing the vocabulary's validity through its precision in human evaluation of vocalization completeness and its recall by determining the coverage rate within dog vocalization sequences. Our analysis revealed acoustic uniformity in the segments of identical phoneme ngrams across various dogs, demonstrating that these ``words'' comprehensively encompass the entirety of dog vocalization sentences.

Our contributions lie in three aspects: 
\begin{enumerate}
\item We developed a robust processing pipeline for dog vocalizations, achieving high accuracy in phoneme labeling.
\item We design a popularity score to measure the probability of a dog phoneme ngram being a dog word, and obtain a vocabulary of dog phonemes without repetition. These words are uttered by multiple dogs and exhibit significant consistency.
\item We developed a web-based dog vocalization labeling system that can analyze and label dog audio uploaded by users, and highlight phoneme ngrams present in the vocabulary, laying an important foundation for further research on dog language understanding.
\end{enumerate}

\section{Method}
\label{sec:method}

\begin{figure}[th]
    \centering
    \scalebox{0.35}{\includegraphics{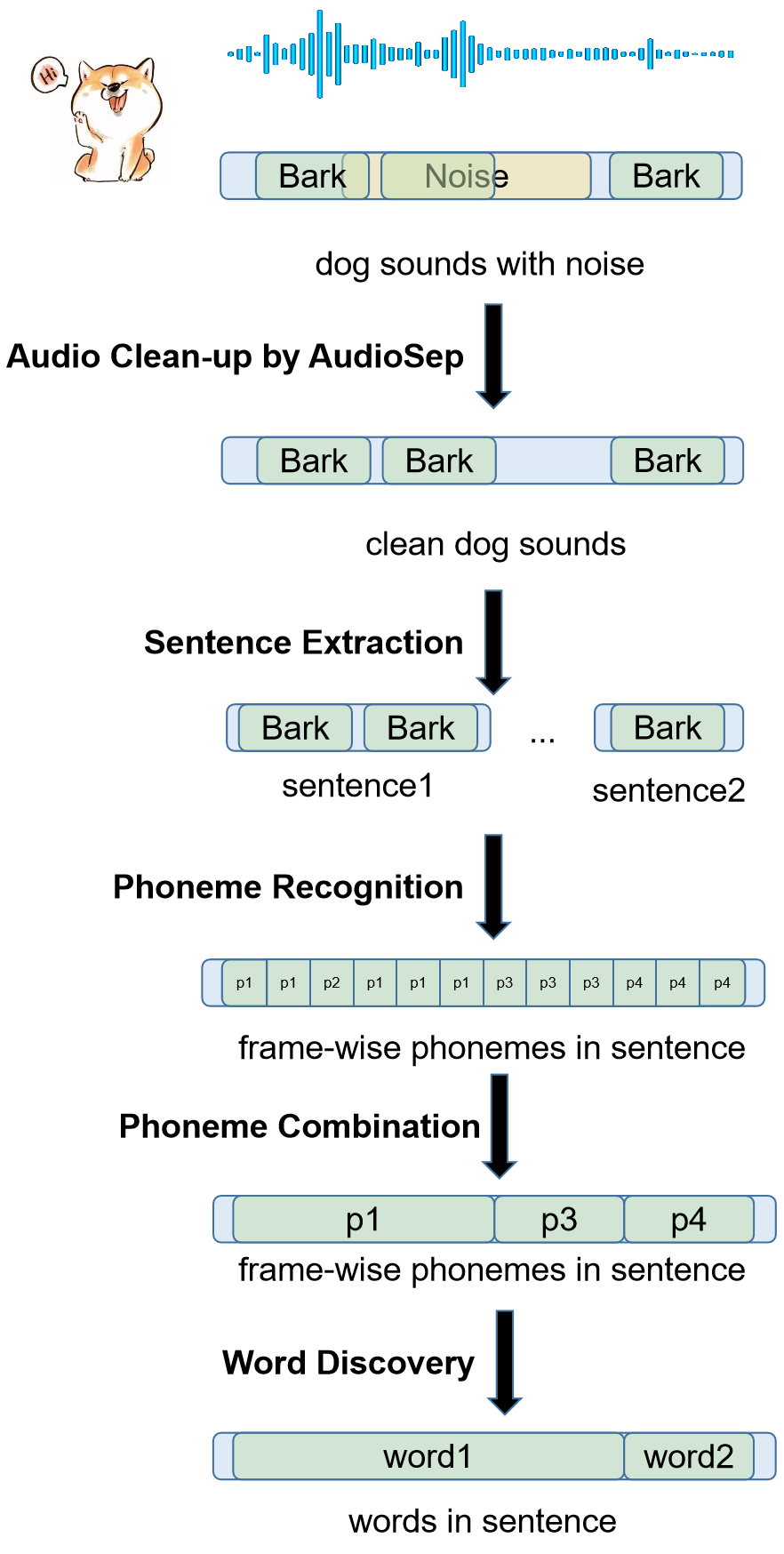}}
    \caption{Full pipeline from data processing to word discovery.
}
    \label{fig:Workflow}
\end{figure}

In this section, we detail our workflow\footnote{The code and a live demo are available at \url{https://anonymous.4open.science/r/Dog_Script-CDB0}}(\figref{fig:Workflow}) including audio clean-up by AudioSep, sentence extracting, phoneme recognition, phoneme combination, and word discovery. 

\subsection{Audio Clean-up by AudioSep}

A mixture of dog sounds and noises will inevitably occur due to the use of videos from public social media. These noises can be background music edited in by the video uploader, human speeches, toy noises, etc. We expect a cleaner dataset, so we need to separate dog sounds from all mixed audios. In this work, we use AudioSep~\citep{liu2023separate}
, a foundation model for open-domain audio source separation with natural language queries. The AudioSep is pre-trained on large-scale multimodal datasets, including AudioSet~\citep{gemmeke2017audio} dataset, VGGSound~\citep{chen2020vggsound} dataset, AudioCaps~\citep{kim2019audiocaps} dataset, etc. We apply AudioSep, using ``Dog'' as the input text query, to separate dog sounds from all audio. If it contains long audios that prevent AudioSep from running, you can start by cutting shorter audio clips with low-threshold~(less than 0.05) PANNs.

To verify the effectiveness of AudioSep and to confirm that it did not have a significant impact on sound quality, we manually labeled 1467 seconds~(1137 seconds for train and 330 seconds for test) audios of dog barking to fine-tune the PANNs~\citep{kong2020panns}. The result~(\tabref{tab:audioSep_res}) shows that AudioSep can effectively reduce noise interference without significantly impacting the quality of dog sounds.

\begin{table}[h]
\centering
\begin{tabular}{lcc}
\hline
\textbf{Train Data} & \textbf{Test Data} & \textbf{F1 Score}\\
\hline
~~~~~~~~~- & - & 0.6916 \\
\hline
~~~~~~~~~\Checkmark{} & - & 0.6797 \\
\hline
~~~~~~~~~- & \Checkmark{} & \pmb{0.7755} \\
\hline
~~~~~~~~~\Checkmark{} & \Checkmark{} & 0.7709 \\
\hline
\end{tabular}
\caption{The result of PANNs. Those with a check mark are using AudioSep.}
\label{tab:audioSep_res}
\end{table}

\subsection{Sentence Extracting}
After the separation of dog sounds, the 
audio will contain mostly barking, silence, and a small amount of noise that can't be separated. To focus on the part of dog sounds, we need to eliminate as much interference as possible from remaining silences and noises. To extract the part of dog sounds, we use a similar approach to \citet{huang2023transcribing}. We first apply PANNs~\citep{kong2020panns} trained on the large-scale AudioSet dataset to extract clips of dog sounds. Then, consecutive clips less than one second apart combine to form a dog ``sentence''. With AudioSep, we can get more and cleaner sentences than \citet{huang2023transcribing}.

To further eliminate noise, we use PANNs for audio tagging. Sentences that meet one of the following conditions will be filtered: (1) ``Dog'' label score less than 0.1. (2) Presence of labels not related to the dog with a score greater than 0.1.

\subsection{Phoneme Recognition}

Dog language is unfamiliar territory for humans. The lexicon, length, and other information of dog sound units are unknown, making it can not easily delineate sound units in a dog sentence in the same way that we delineate words in a human sentence. Directly using a priori knowledge of human language or the way to split human language units like \citet{huang2023transcribing} to understand dog language may be inapplicable. Self-supervised approaches for speech representation learning may be a more appropriate method for discovering the sound units of dog language. In this work, we use HuBERT, a self-supervised approach that learns a combined acoustic and language model over continuous inputs, to understand dog sound units. In \citeposs{hagiwara2023aves} work, Hubert has demonstrated a good ability to represent animal vocalization. So we use dog ``sentences'' to pre-train HuBERT. 

\begin{figure}[h]
    \centering
    \scalebox{0.35}{\includegraphics{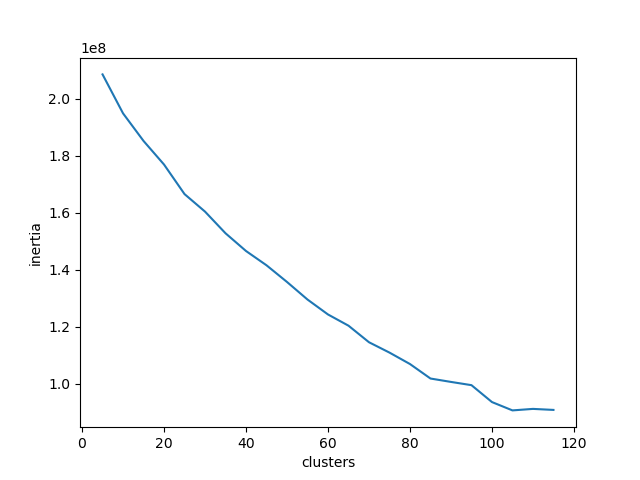}}
    \caption{Inertia under different clusters. 50 is a suitable clusters. This is the basis for our choice of 50 clusters in the third K-Means model. The same method was used for the others.}
    \label{fig:clusters}
\end{figure}

For HuBERT Pretraining, we used 54 clusters at the first stage and 100 clusters at the second stage, a learning rate of 0.0001, and 100k training steps at the first stage and 109k training steps at the second stage. Then we used features, which are from the 12th transformer layer of the second-stage model, to train a K-Means model with 50 clusters. The clusters are determined based on the sum of distances from each sample to its cluster center for different clusters (\figref{fig:clusters}).

\subsection{Phoneme Combination}
After K-means clustering the features extracted from the approximately 20ms audio frames using HuBERT, we obtained a transcript with labels for each dog sound sentence. However, interestingly, the duration of a phoneme is often longer than 20ms. With the help of the excellent context recognition ability of the Transformer-based model, a large number of labels appear continuously in the transcript. After observing it, we found that there are still some noise segments in the transcript, such as a transcript containing the sequence: $[27, 27, 27, 5, 27, 27]$. After manual review of the original video, we found that the segment contains some faint noise. In order to obtain a purer transcript, we designed a phoneme combination algorithm.


The algorithm uses double pointers to probe whether there are segments with the same label on both sides of the label. tolerance refers to the maximum length that can be assimilated by the labels on both sides. For example, when $tolerance=1$, the transcript segment $[a, a, b, b, a, a]$ cannot be optimized to $[a, a, a, a, a, a]$. After global phone combination optimization, we obtain the mean and standard deviation of the length of each phoneme.

\subsection{Word Discovery}

Once we obtained accurate labels for dog vocalization phonemes, we explored how to determine the ``word'' of a dog based on the current transcription.

\subsubsection{Enumeration}

To obtain a accurate ``vocabulary'', our first step is to list all possible words using the N-gram algorithm, which becomes the candidate vocabulary.
\begin{algorithm}
\caption{Enumeration Algorithm}\label{alg:cap}
\begin{algorithmic}
\Require{$S \gets [s_1, s_2, s_3, ...]$}
\Require{$n$}
\Ensure{$ngrams$}
\State $ngrams \gets [ ]$
\State $i \gets 0$
\For{\texttt{seq in S}}
\State $l \gets len(seq)$
\State $i \gets 0$
\While{$i \leq l - n + 1$}
\State $ngram\gets seq[i:i+n]$
\State $i \gets i+1$
\State $ngrams$.append($ngram$)
\EndWhile
\EndFor
\end{algorithmic}
\end{algorithm}

\subsubsection{Scoring}

We believe that a word must satisfy the condition that it appears frequently enough in the transcription information to indicate its repeatability, and at the same time, we must ensure that the word is uttered by more than one dog to determine its universality. To this end, we designed a popularity score, the formula is as follows
$$Ps(gram^n_i) = f(gram^n_i) * \delta(gram^n_i)$$
, where $Ps(\cdot)$ is the popularity score function, $f(\cdot)$ is a function to calculate the frequency of this gram, $\delta(\cdot)$ is a function to calculate the diversity of this word, $gram^n_i$ is a unique n-gram, $n$ is the number of frames, $i$ indicates the $i_{th}$ unique ngram. The formula of $f$ and $\delta$ is as follows: 
\begin{eqnarray*}
f(gram^n_i) &=& \frac{\left | gram^n_i\right |}{\sum_{i}\left |gram^n_i\right |)}\\
\delta(gram^n_i) &=& \left |\{x\in D:x\ contains\ gram^n_i\}\right|,
\end{eqnarray*}
where $D$ is a set of different dogs in training dataset, $\left |\cdot \right |$ means to get the number of a set. 
The higher the popularity score of an ngram, the stronger the universality of the ngram, and its sound is uttered more times by more dogs.

\subsubsection{Filtering}

Once we have obtained the popularity score of all ngrams, the next question is how to filter them into the vocabulary. The first condition for an ngram to be called a ``word'' is its completeness. Inevitably, our ngrams with high popularity scores must contain incomplete ``words'', which are often part of a longer ngram. However, if we do not limit the selection of longer ngrams, our ``words'' will change from a single word to 2 or even a sentence. However, setting a threshold for the popularity score can help us avoid this problem. 

To this end, we set a hyperparameter \textbf{threshold} to ensure that the popularity score of ngrams included in the vocabulary is large enough, and start from the longest ngram and make sure that all words shorter than the ngram will not be part of the selected ngram. This ensures that the words in the vocabulary have complete attributes as much as possible. The selection of this hyperparameter can be done through recall, which will be discussed in \secref{sec:exp}.

\section{Experiments}
\label{sec:exp}

\subsection{Dataset}

We use a similar approach to \citet{wang2023towards} to download videos. The dataset contains 27,775 sentences(more than 20 hours) from 13,143 dog related videos.

\subsection{Phoneme Evaluation}

\paragraph{Evaluation Protocol}

A successfully classified phone should possess the following two properties: 1) phones with the same label (phoneme) should sound very similarly; 
and 2) phones with different labels are sounds that are clearly distinct. 
To verify the reliability of the dog vocalization phonemes obtained in 
\secref{sec:method}, we design a phoneme comparison experiment.
The dog vocalization sentences in the test dataset were processed using 
the operations described in \secref{sec:method} to obtain phoneme label 
sequences. Each type of label was randomly sampled and intercepted. 
To improve the accuracy of the identification by human beings, 
we intercepted phones of at least 5 frames
(with a length greater than 0.1 seconds) as samples for the accuracy test. The frames are from HuBERT(about 20ms a frame).

\paragraph{Annotator Demographics}

The testers are two college students majoring in engineering who love small animals and participated in the experiment as volunteers.

The testers will listen to several pairs of audio segments, including segments with the same label and segments with different labels. The testers are required to use their prior knowledge to judge whether the pair of audio segments belongs to the same category.

\begin{table}
\centering
\small
\begin{tabular}{lll}
\hline
\textbf{Type} & \textbf{Labels}\\
\hline
\verb|Dog| & 1, 7, 9, 10, 11, 12, 14, 16, 17, 20-24, 26-34, \\
& 37, 40-44, 46-50 \\
\verb|Noise| & 2, 3, 4, 5, 6, 8, 13, 15, 18, 19, 25, 35, 36, \\
& 38, 39, 45 \\\hline
\end{tabular}
\caption{Dog phonemes vs noice phonemes}
\label{tab:phoneclassification}
\end{table}

\paragraph{Metrics}

In the first round of testing, two testers randomly selected 30 labels from all labels and randomly sampled 8 pairs of identical audio segments and 7 pairs of different audio segments in each label, for a total of 225 pairs of audio segments. The accuracy is shown in \tabref{tab:phonetestresult}.

During the test, we found that almost $1/3$ of the labels corresponded to noise segments. To this end, two testers conducted a noise-dog sound discrimination experiment on all 50 labels. The experiment randomly selected 5 samples from all 51 labels. The two testers first judged whether the segment was noise from the sound features. If it was impossible to judge from the audio features alone, the testers would refer to the corresponding video for judgment. The judgment results of the two testers are shown in \tabref{tab:phoneclassification}.

In order to ensure the accuracy of the consistency test of dog sound labels, after excluding noise labels, we finally sampled 30 phoneme labels randomly in dog vocalization labels. Each label has 10 test groups, including five identical label pairs and five different label pairs, for a total of 300 pairs of audio segments. The accuracy is shown in \tabref{tab:phonetestresult}.

\begin{figure}[th]
\scalebox{0.1}{\includegraphics{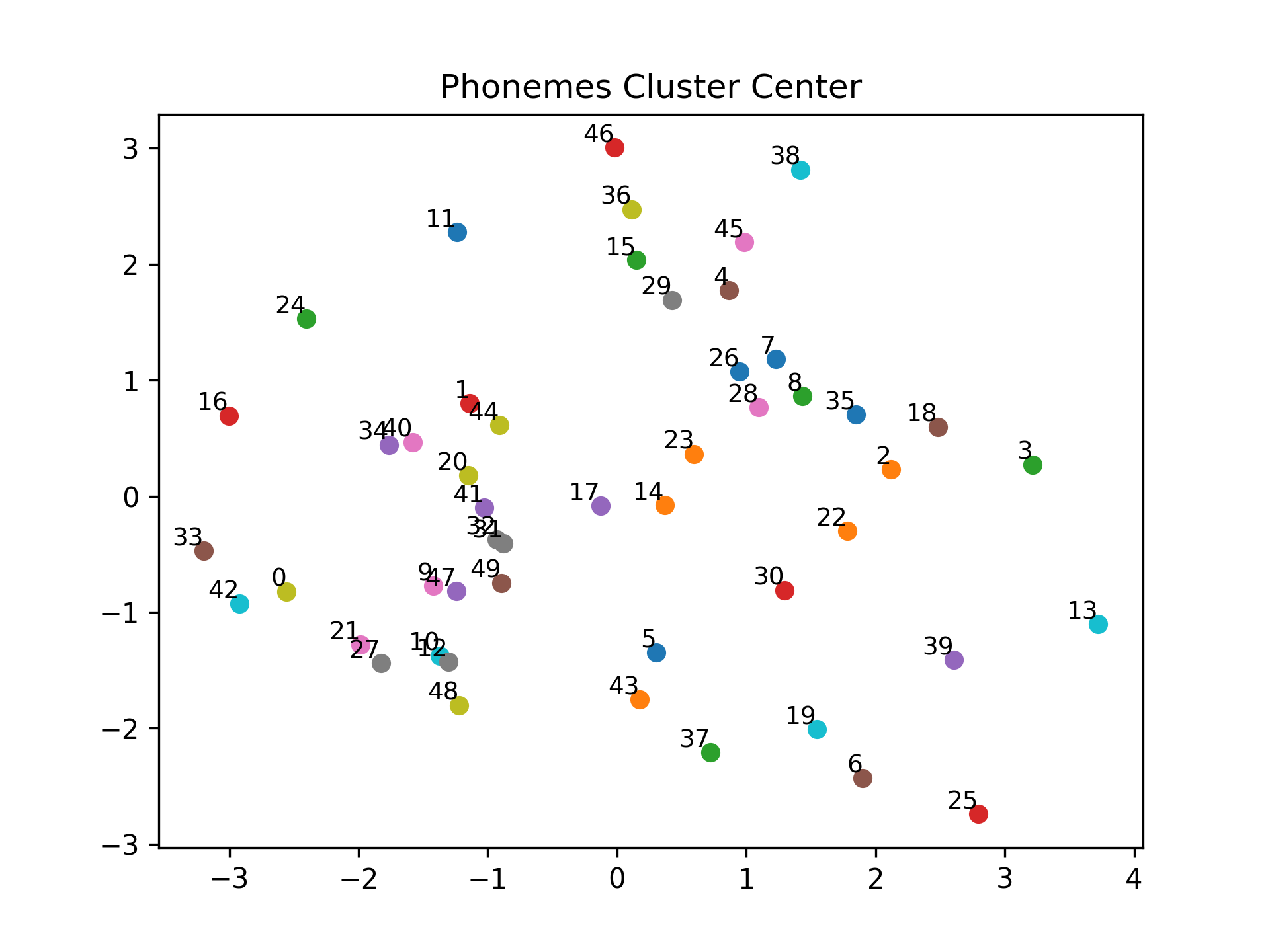}}
\caption{2-D Visualization of 50 phonemes from HuBERT.}
\label{fig:cluster}
\end{figure}

From the test results, we can find that the test results containing only dog sound labels are better than the test data containing noise. At the same time, the testers' judgments on the consistency of the sound are relatively consistent, and the agreement rate is above 80\% on the individual dog labels. According to the feedback from the testers, due to the large number of noise segment types, different types of noise are often classified into the same label within a limited number of labels. The clustering center point results of the audio vector representation of the 50 labels are shown in \figref{fig:cluster}. Due to the model's focus on the context based on transform, noise labels often exist around the dog sound labels, making it difficult for testers to distinguish.



\begin{table}[th]
\centering
\small
\begin{tabular}{lcc}
\hline
\textbf{Tester} & \textbf{dog voice label} & \textbf{total label}\\
\hline
\verb|Tester 1| & 72.0 & 66.67 \\
\verb|Tester 2| & 70.5 & 64.89 \\
\verb|Agreement| & 80.5 & 74.22 \\\hline
\end{tabular}
\caption{Accuracy and agreement result on testing the reliability of phonemes discovery}
\label{tab:phonetestresult}
\end{table}


Overall, the phoneme labels obtained are consistent with the phonemes, 
and their classification is relatively accurate. This provides us with the 
possibility to further explore how to obtain lexical discoveries of 
dog vocalizations.

\subsection{Lexical Evaluation}

\paragraph{Evaluation Protocol}

After the combination algorithm we mentioned in \secref{sec:method}, the dog vocalization sentence will be transcribed into a list composed of phoneme labels. Will it contain combinations of labels with fixed patterns, such as words? 
As one of the constituent elements of language, words should have the 
two characteristics of 1) frequently appearing in voice samples and 2) being 
able to be uttered by many different individuals. 
To obtain such a ``vocabulary'', we followed the two basic characteristics of words and obtained a vocabulary composed of bigrams, trigrams, and up to six grams, according to the process described in \secref{sec:method}.

\begin{figure*}[h]
\centering
\scalebox{0.3}{\includegraphics{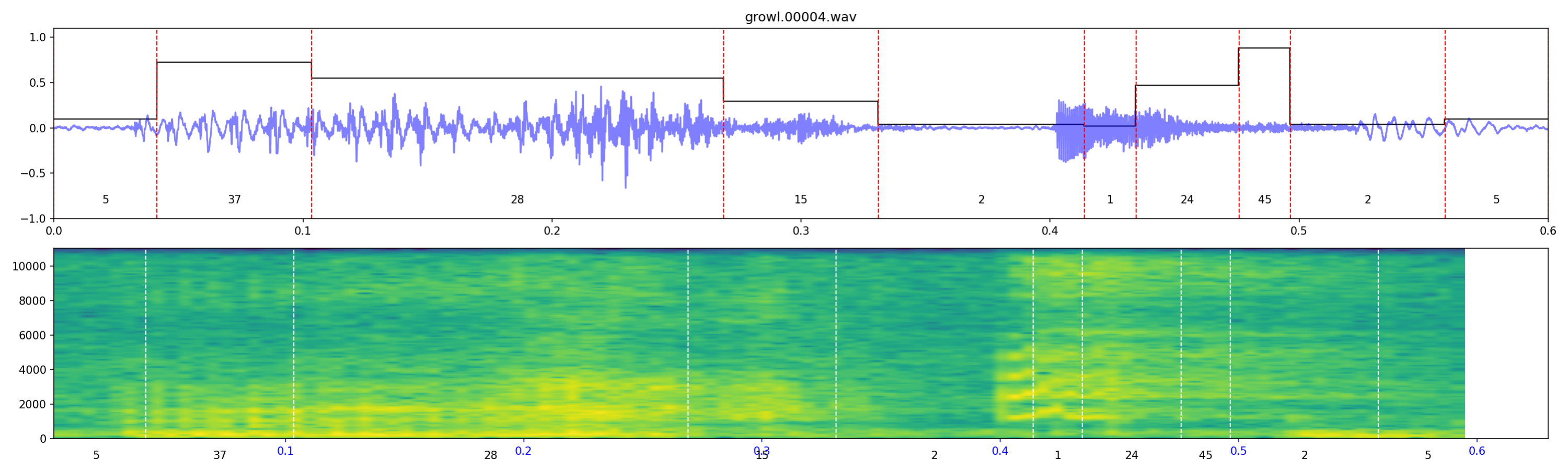}}
\caption{Segmentation and Phoneme Labelling Result of A Growl Sound. }
\label{fig:cwc}
\end{figure*}

\begin{figure}[th]
\centering
\scalebox{0.3}{\includegraphics{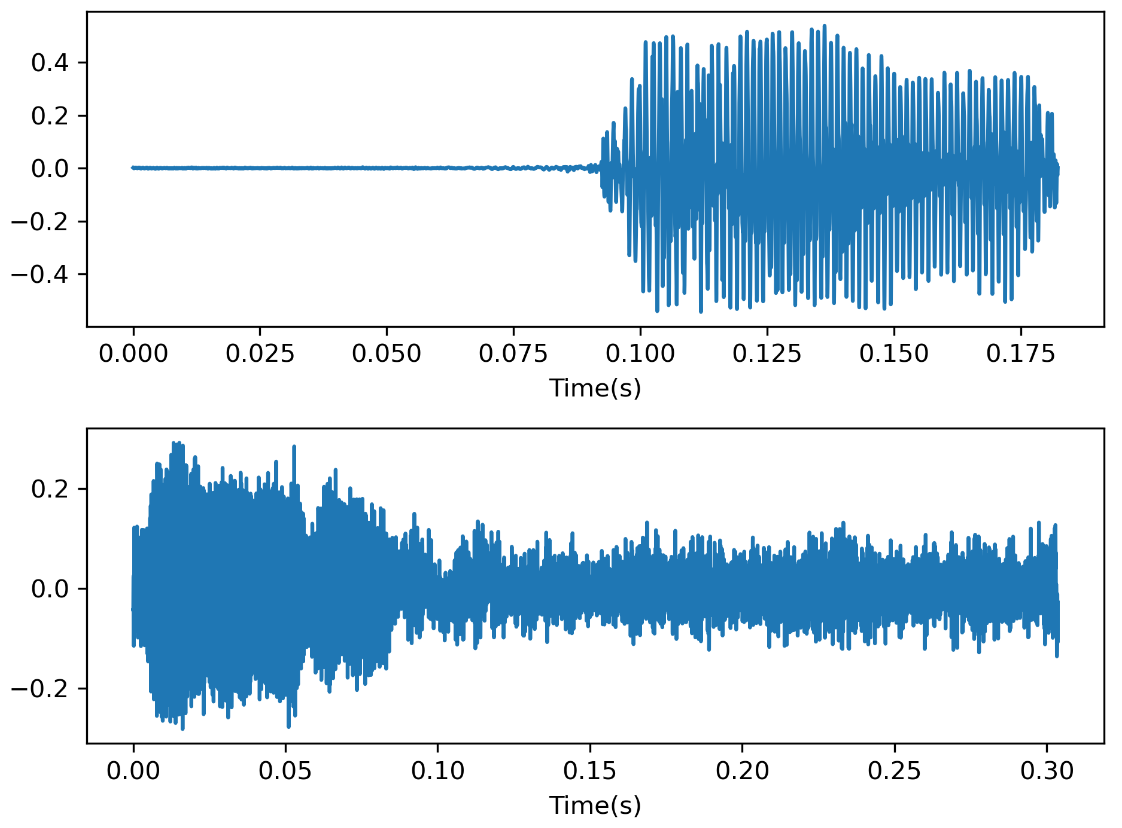}}
\caption{Top: waveform of word (18, 35, 40, 33); Bottom: waveform of word (24, 26)}
\label{fig:full}
\end{figure}

Due to the accuracy of the phoneme classification labels, we shift our attention from the consistency of audio segments in the same ngram to the integrity of words in the ``vocabulary''. A word usually contains only one continuous sound, which can be reflected in the energy diagram of the audio. As shown in Figure \ref{fig:full}, whether it is based on the characteristics of the audio or from the observation angle of the energy diagram, we believe that the audio at the top is a complete pronunciation, while the audio at the bottom is not. 

In addition to measuring whether an ngram is a complete word and whether the same ngram has acoustic consistency, we also need to determine that the words in the ``vocabulary'' have a certain degree of universality, i.e., they are contained in a wide range of individual voices. 


\paragraph{Metrics}

Two testers score 100 randomly selected samples from the ``vocabulary'', with 1 points indicating that the sample is a complete word, 0 points indicating that the sample is not a complete word. The test results is 67\%. Inevitably, due to the shorter length of Bigram, there are more of them under the same threshold; from the feedback of the testers, although some Bigram is represented as the nasal sound of a dog, there are still some incomplete audio. We speculate that this is mainly because the universality of this segment is stronger, and there are many kinds of phonemes that can be matched before and after, which leads to its higher $\delta$ value, thus increasing the value of its popularity score.

To measure the recall of the ``vocabulary'', we design two metrics, phoneme coverage and phone coverage to evaluate whether the ``vocabulary'' is universal. Phoneme coverage refers to the percentage of the number of phonemes from all words in the ``vocabulary'' to the total number of phonemes in all sentences; phone coverage refers to the percentage of the total audio duration of all words from the vocabulary to the total audio duration of all sentences.

\begin{figure}[th]
    \centering
    \scalebox{0.1}{\includegraphics{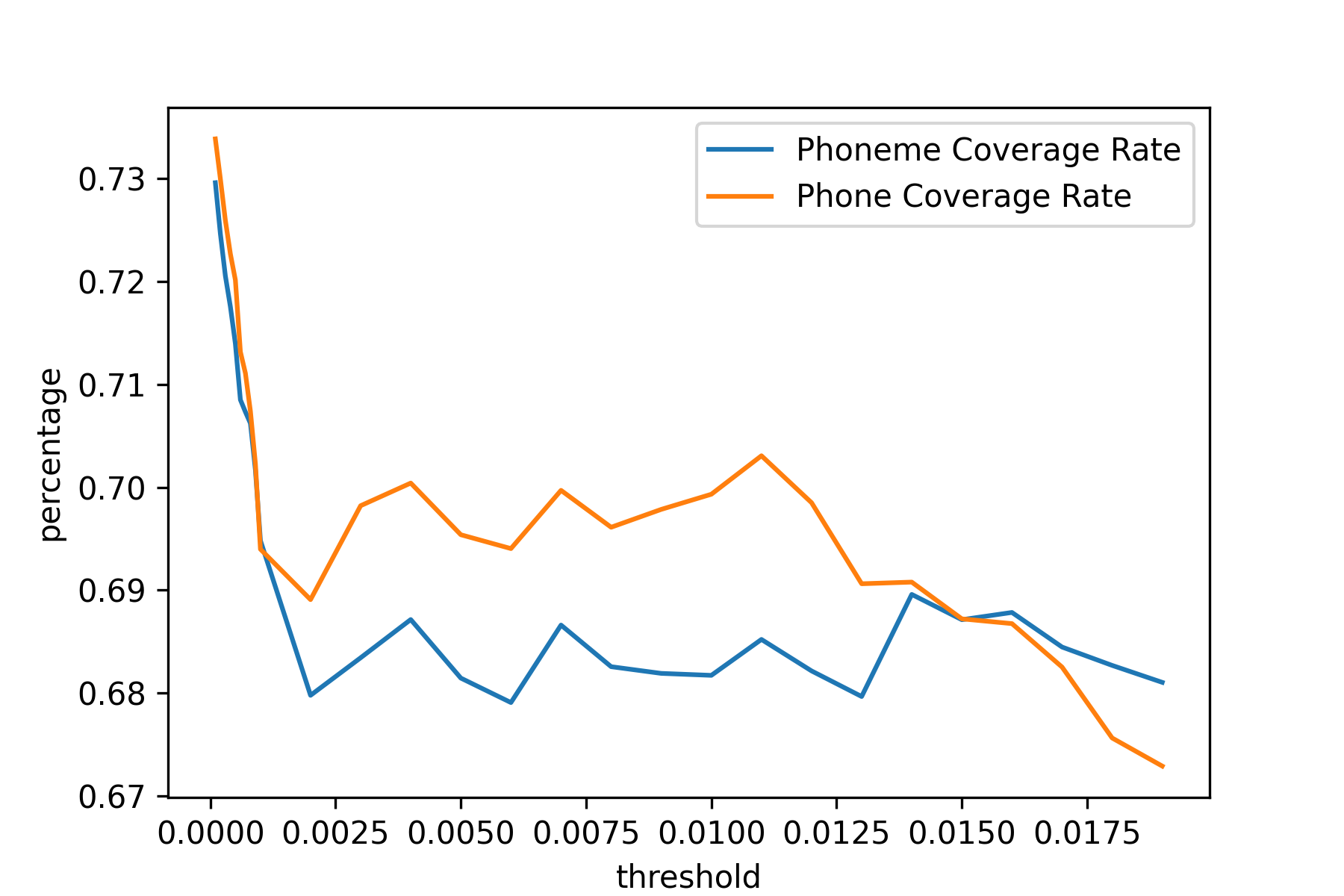}}
    \caption{Phoneme Coverage and Phone Coverage on different threshold}
    \label{fig:wdc}
\end{figure}

From another perspectives, these two metrics can also help us determine the value of the threshold used to divide the ``vocabulary''. The lower the threshold value, the more words in the vocabulary, and these additional words in the vocabulary, and these additional words may also have their own meanings. However, in this work, we want to ensure that each word in the ``vocabulary'' has the highest probability of having a meaning on the basis of ensuring macro word coverage. The coverage rate results for different thresholds are shown in Figure \ref{fig:wdc}.


\section{Demonstration}

A web-based dog vocalization labeling system can intuitively show us its energy diagram and spectrogram, and we can also refer to this intuitive representation to verify whether the phonemes with the same label are the same. The system can be used to further explore the question of whether dog vocalization has certain meanings. The system is mainly divided into three parts. The first part is the introduction, which introduces how the back-end implements the dog vocalization labeling system and provides a clustering center distribution diagram of the clustering results of the model reduced to two dimensions by PCA, and how to use the other two functions.

\begin{figure*}[h]
    \centering
    \scalebox{0.4}{\includegraphics{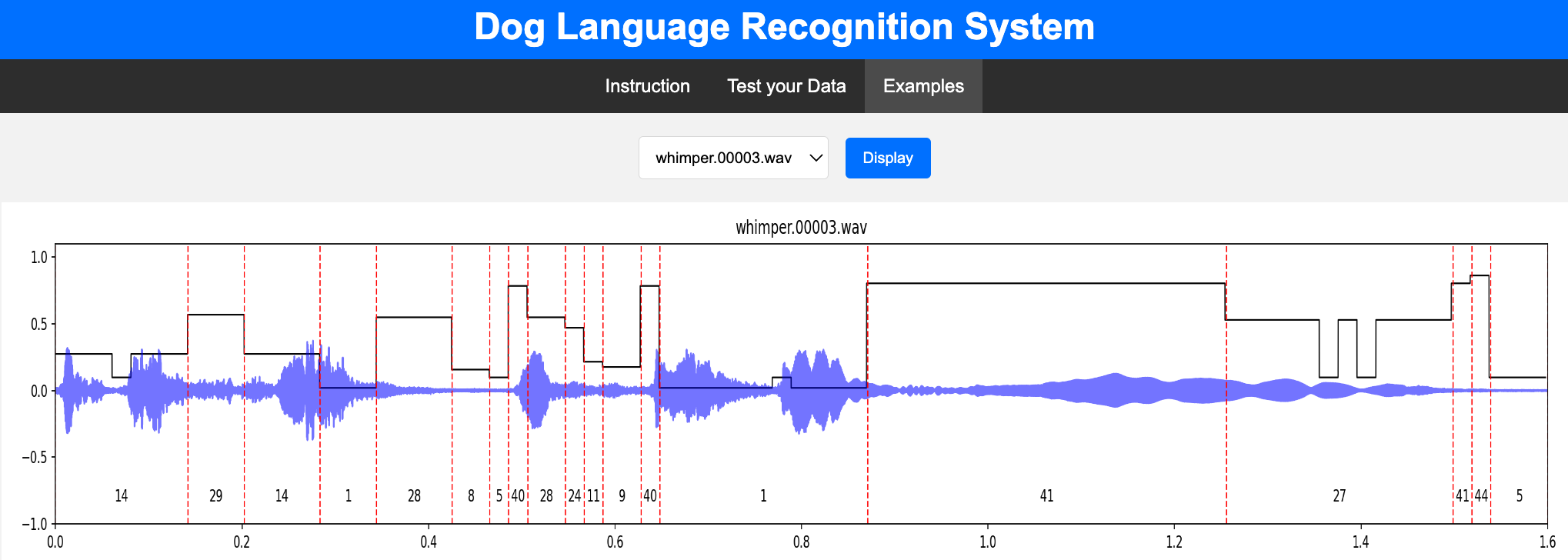}}
    \caption{Examples interface of Dog Language Recognition System.}
    \label{fig:demo1}
\end{figure*}

The second part provides users with the function of transcribing their own recorded dog vocalization, and displays their energy diagram, spectrogram, and corresponding label sequence, the interface of main page is shown as \figref{fig:demo1}. The original label sequence that has not undergone the Phone Combination operation will also be represented in the energy diagram, which can be compared with its final label and with the audio to measure the performance of the phone combination algorithm. It also supports listening to the sound of each label's corresponding phoneme alone, and the audio clip of the ``word'' in the vocabulary we found. In the next step, this page will also support the playback of video clips corresponding to the audio clips, so as to further study the meaning of each word.

The system also provides a sample display function. Each word in the vocabulary and its sound are displayed on this page, and 20 dog language sentences from the training dataset are randomly selected so that users who do not have dog vocalization samples can experience the functions of the system.

\section{Related Work}

To decode a dog's language, it is necessary to analyze its basic sound units, linguistic structure, lexicon, meaning, etc. 
The past few years have seen a surge of interest in using machine learning (ML) methods for studying the behavior of nonhuman animals~\citep{rutz2023using}. 
Much of the past work has mostly focused on dog behavior~\citep{ide2021rescue,ehsani2018let} and the meaning of dog sounds~\citep{molnar2008classification,hantke2018my,larranaga2015comparing,hantke2018my,pongracz2006acoustic}. 
Most of them only classify the audio of the dog sounds in multiple categories, including activities, contexts, emotions, ages, etc.
They did not study the sound units of the dog's language.

The above work likewise illustrates the existence of multiple distinct sound units in dog language. Many of the species that appear to use only a handful of basic call types may turn out to possess rich vocal repertoires~\citep{rutz2023using}. Many works demonstrate the diversity of animal sounds~\citep{paladini2020bark,robbins2000vocal,bermant2019deep}. \citet{huang2023transcribing} and \citet{wang2023towards} did a fine-grained study of dog sound units. They directly using a priori knowledge of human language may be inapplicable.
We use HuBERT~\citep{hsu2021hubert}, a self-supervised approach, to find the sound units of dog language, and future research will find the meaning of sound units.

\section{Conclusion}

In this paper, we present a method with HuBERT to explore canine sound units. In contrast to previous work, this approach uses a self-supervised method, instead of relying on human language knowledge, to explore sound units in canine language. This is better suited for exploring fine-grained sound units in canine language.

We used this method to obtain a vocabulary of dog phonemes without repetition. These vocabulary words are uttered by multiple dogs and exhibit significant consistency. Future research can detect other information about the dog, such as behavior, mood, location, etc., to explore the meaning carried by these vocabulary words. We developed a web-based dog vocalization labeling system. Users can upload a dog audio and easily access the words in the vocabulary we found.

\section*{Limitation}

The discovery of dog sound units heavily relies on the quality of the dataset. 
Even though we have implemented multiple measures to enhance the dataset's quality, noise may still be present due to various factors, including the recording equipment, background noise, or added noise by video uploaders. 
Further research may focus on finetune PANNs or other sound event detection models to acquire datasets of higher quality.


\bibliography{references}

\begin{thebibliography}{20}
\expandafter\ifx\csname natexlab\endcsname\relax\def\natexlab#1{#1}\fi

\bibitem[{Bermant et~al.(2019)Bermant, Bronstein, Wood, Gero, and Gruber}]{bermant2019deep}
Peter~C Bermant, Michael~M Bronstein, Robert~J Wood, Shane Gero, and David~F Gruber. 2019.
\newblock Deep machine learning techniques for the detection and classification of sperm whale bioacoustics.
\newblock \emph{Scientific reports}, 9(1):12588.

\bibitem[{Chen et~al.(2020)Chen, Xie, Vedaldi, and Zisserman}]{chen2020vggsound}
Honglie Chen, Weidi Xie, Andrea Vedaldi, and Andrew Zisserman. 2020.
\newblock Vggsound: A large-scale audio-visual dataset.
\newblock In \emph{ICASSP 2020-2020 IEEE International Conference on Acoustics, Speech and Signal Processing (ICASSP)}, pages 721--725. IEEE.

\bibitem[{Ehsani et~al.(2018)Ehsani, Bagherinezhad, Redmon, Mottaghi, and Farhadi}]{ehsani2018let}
Kiana Ehsani, Hessam Bagherinezhad, Joseph Redmon, Roozbeh Mottaghi, and Ali Farhadi. 2018.
\newblock Who let the dogs out? modeling dog behavior from visual data.
\newblock In \emph{Proceedings of the IEEE Conference on Computer Vision and Pattern Recognition}, pages 4051--4060.

\bibitem[{Gemmeke et~al.(2017)Gemmeke, Ellis, Freedman, Jansen, Lawrence, Moore, Plakal, and Ritter}]{gemmeke2017audio}
Jort~F Gemmeke, Daniel~PW Ellis, Dylan Freedman, Aren Jansen, Wade Lawrence, R~Channing Moore, Manoj Plakal, and Marvin Ritter. 2017.
\newblock Audio set: An ontology and human-labeled dataset for audio events.
\newblock In \emph{2017 IEEE international conference on acoustics, speech and signal processing (ICASSP)}, pages 776--780. IEEE.

\bibitem[{Hagiwara(2023)}]{hagiwara2023aves}
Masato Hagiwara. 2023.
\newblock Aves: Animal vocalization encoder based on self-supervision.
\newblock In \emph{ICASSP 2023-2023 IEEE International Conference on Acoustics, Speech and Signal Processing (ICASSP)}, pages 1--5. IEEE.

\bibitem[{Hantke et~al.(2018)Hantke, Cummins, and Schuller}]{hantke2018my}
Simone Hantke, Nicholas Cummins, and Bjorn Schuller. 2018.
\newblock What is my dog trying to tell me? the automatic recognition of the context and perceived emotion of dog barks.
\newblock In \emph{2018 IEEE International Conference on Acoustics, Speech and Signal Processing (ICASSP)}, pages 5134--5138. IEEE.

\bibitem[{Holdcroft(1991)}]{holdcroft1991saussure}
David Holdcroft. 1991.
\newblock \emph{Saussure: signs, system and arbitrariness}.
\newblock Cambridge University Press.

\bibitem[{Hsu et~al.(2021)Hsu, Bolte, Tsai, Lakhotia, Salakhutdinov, and Mohamed}]{hsu2021hubert}
Wei-Ning Hsu, Benjamin Bolte, Yao-Hung~Hubert Tsai, Kushal Lakhotia, Ruslan Salakhutdinov, and Abdelrahman Mohamed. 2021.
\newblock Hubert: Self-supervised speech representation learning by masked prediction of hidden units.
\newblock \emph{IEEE/ACM Transactions on Audio, Speech, and Language Processing}, 29:3451--3460.

\bibitem[{Huang et~al.(2023)Huang, Zhang, Wu, and Zhu}]{huang2023transcribing}
Jieyi Huang, Chunhao Zhang, Mengyue Wu, and Kenny Zhu. 2023.
\newblock Transcribing vocal communications of domestic shiba lnu dogs.
\newblock In \emph{Findings of the Association for Computational Linguistics: ACL 2023}, pages 13819--13832.

\bibitem[{Ide et~al.(2021)Ide, Araki, Hamada, Ohno, and Yanai}]{ide2021rescue}
Yuta Ide, Tsuyohito Araki, Ryunosuke Hamada, Kazunori Ohno, and Keiji Yanai. 2021.
\newblock Rescue dog action recognition by integrating ego-centric video, sound and sensor information.
\newblock In \emph{Pattern Recognition. ICPR International Workshops and Challenges: Virtual Event, January 10--15, 2021, Proceedings, Part III}, pages 321--333. Springer.

\bibitem[{Kim et~al.(2019)Kim, Kim, Lee, and Kim}]{kim2019audiocaps}
Chris~Dongjoo Kim, Byeongchang Kim, Hyunmin Lee, and Gunhee Kim. 2019.
\newblock Audiocaps: Generating captions for audios in the wild.
\newblock In \emph{Proceedings of the 2019 Conference of the North American Chapter of the Association for Computational Linguistics: Human Language Technologies, Volume 1 (Long and Short Papers)}, pages 119--132.

\bibitem[{Kong et~al.(2020)Kong, Cao, Iqbal, Wang, Wang, and Plumbley}]{kong2020panns}
Qiuqiang Kong, Yin Cao, Turab Iqbal, Yuxuan Wang, Wenwu Wang, and Mark~D Plumbley. 2020.
\newblock Panns: Large-scale pretrained audio neural networks for audio pattern recognition.
\newblock \emph{IEEE/ACM Transactions on Audio, Speech, and Language Processing}, 28:2880--2894.

\bibitem[{Larranaga et~al.(2015)Larranaga, Bielza, Pongr{\'a}cz, Farag{\'o}, B{\'a}lint, and Larranaga}]{larranaga2015comparing}
Ana Larranaga, Concha Bielza, P{\'e}ter Pongr{\'a}cz, Tam{\'a}s Farag{\'o}, Anna B{\'a}lint, and Pedro Larranaga. 2015.
\newblock Comparing supervised learning methods for classifying sex, age, context and individual mudi dogs from barking.
\newblock \emph{Animal cognition}, 18(2):405--421.

\bibitem[{Liu et~al.(2023)Liu, Kong, Zhao, Liu, Yuan, Liu, Xia, Wang, Plumbley, and Wang}]{liu2023separate}
Xubo Liu, Qiuqiang Kong, Yan Zhao, Haohe Liu, Yi~Yuan, Yuzhuo Liu, Rui Xia, Yuxuan Wang, Mark~D Plumbley, and Wenwu Wang. 2023.
\newblock Separate anything you describe.
\newblock \emph{arXiv preprint arXiv:2308.05037}.

\bibitem[{Moln{\'a}r et~al.(2008)Moln{\'a}r, Kaplan, Roy, Pachet, Pongr{\'a}cz, D{\'o}ka, and Mikl{\'o}si}]{molnar2008classification}
Csaba Moln{\'a}r, Fr{\'e}d{\'e}ric Kaplan, Pierre Roy, Fran{\c{c}}ois Pachet, P{\'e}ter Pongr{\'a}cz, Antal D{\'o}ka, and {\'A}d{\'a}m Mikl{\'o}si. 2008.
\newblock Classification of dog barks: a machine learning approach.
\newblock \emph{Animal Cognition}, 11:389--400.

\bibitem[{Paladini(2020)}]{paladini2020bark}
Aleida Paladini. 2020.
\newblock The bark and its meanings in inter and intra-specific language.
\newblock \emph{Dog behavior}, 6(1):21--30.

\bibitem[{Pongr{\'a}cz et~al.(2006)Pongr{\'a}cz, Moln{\'a}r, and Miklosi}]{pongracz2006acoustic}
P{\'e}ter Pongr{\'a}cz, Csaba Moln{\'a}r, and Adam Miklosi. 2006.
\newblock Acoustic parameters of dog barks carry emotional information for humans.
\newblock \emph{Applied Animal Behaviour Science}, 100(3-4):228--240.

\bibitem[{Robbins(2000)}]{robbins2000vocal}
Robert~L Robbins. 2000.
\newblock Vocal communication in free-ranging african wild dogs (lycaon pictus).
\newblock \emph{Behaviour}, pages 1271--1298.

\bibitem[{Rutz et~al.(2023)Rutz, Bronstein, Raskin, Vernes, Zacarian, and Blasi}]{rutz2023using}
Christian Rutz, Michael Bronstein, Aza Raskin, Sonja~C Vernes, Katherine Zacarian, and Dami{\'a}n~E Blasi. 2023.
\newblock Using machine learning to decode animal communication.
\newblock \emph{Science}, 381(6654):152--155.

\bibitem[{Wang et~al.(2023)Wang, Zhang, Huang, Wu, and Zhu}]{wang2023towards}
Yufei Wang, Chunhao Zhang, Jieyi Huang, Mengyue Wu, and Kenny Zhu. 2023.
\newblock Towards lexical analysis of dog vocalizations via online videos.
\newblock \emph{arXiv preprint arXiv:2309.13086}.

\end{thebibliography}

\appendix

\section{Extracting Sentences and HuBERT Pretraining}
\label{sec:appendix_a}

To preserve the full dog ``sentence'', we set the threshold in the cut sentence to:
\begin{equation*}
    threshold = \mathop{min}(0.75 \cdot \mathop{max}(framewise), 0.5)
\end{equation*}
Then increase the duration of each audio clip by 0.5 seconds at the starting and ending points.

For HuBERT Pretraining, we used 54 clusters at the first stage and 100 clusters at the second stage, a learning rate of 0.0001, and 100k training steps at the first stage and 109k training steps at the second stage. Then we used features, which are from the 12th transformer layer of the second-stage model, to train a K-Means model with 50 clusters.

\section{Phonemes Information}

\begin{table}[th]
\centering
\small
\begin{tabular}{lcc}
\hline
\textbf{Phoneme} & \textbf{Mean Length} & \textbf{Variance of Length}\\
\hline
\verb|0| & 0.0569 & 0.0009 \\
\verb|1| & 0.1031 & 0.0085 \\
\verb|2| & 0.0927 & 0.0098 \\
\verb|3| & 0.1224 & 0.0181 \\
\verb|4| & 0.0873 & 0.0063 \\
\verb|5| & 0.0413 & 0.0008 \\
\verb|6| & 0.1033 & 0.0152 \\
\verb|7| & 0.0837 & 0.0054 \\
\verb|8| & 0.0916 & 0.0082 \\
\verb|9| & 0.1253 & 0.0239 \\
\verb|10| & 0.1045 & 0.0097 \\
\verb|11| & 0.0934 & 0.0057 \\
\verb|12| & 0.2063 & 0.0690 \\
\verb|13| & 0.0983 & 0.0104 \\
\verb|14| & 0.1275 & 0.0189 \\
\verb|15| & 0.1070 & 0.0150 \\
\verb|16| & 0.0944 & 0.0058 \\
\verb|17| & 0.1018 & 0.0092 \\
\verb|18| & 0.0737 & 0.0046 \\
\verb|19| & 0.1602 & 0.0279 \\
\verb|20| & 0.0787 & 0.0035 \\
\verb|21| & 0.1685 & 0.0399 \\
\verb|22| & 0.0878 & 0.0073 \\
\verb|23| & 0.0933 & 0.0070 \\
\verb|24| & 0.0960 & 0.0046 \\
\verb|25| & 0.1313 & 0.0213 \\
\verb|26| & 0.1075 & 0.0115 \\
\verb|27| & 0.1387 & 0.0229 \\
\verb|28| & 0.1017 & 0.0082 \\
\verb|29| & 0.0931 & 0.0111 \\
\verb|30| & 0.1018 & 0.0120 \\
\verb|31| & 0.2208 & 0.0786 \\
\verb|32| & 0.2002 & 0.0576 \\
\verb|33| & 0.0655 & 0.0016 \\
\verb|34| & 0.0826 & 0.0102 \\
\verb|35| & 0.0479 & 0.0008 \\
\verb|36| & 0.0689 & 0.0018 \\
\verb|37| & 0.1820 & 0.0526 \\
\verb|38| & 0.0750 & 0.0041 \\
\verb|39| & 0.0947 & 0.0104 \\
\verb|40| & 0.0342 & 0.0002 \\
\verb|41| & 0.1187 & 0.0189 \\
\verb|42| & 0.0766 & 0.0017 \\
\verb|43| & 0.1613 & 0.0312 \\
\verb|44| & 0.1183 & 0.0069 \\
\verb|45| & 0.1223 & 0.0254 \\
\verb|46| & 0.0705 & 0.0018 \\
\verb|47| & 0.1089 & 0.0127 \\
\verb|48| & 0.2405 & 0.0709 \\
\verb|49| & 0.1162 & 0.0211 \\\hline
\end{tabular}
\caption{Mean Length and Variance of Length for each phoneme}
\label{tab:mlvl}
\end{table}

\begin{table*}[h]
\centering
\begin{tabular}{llllll}
\hline
\textbf{2-gram} & \textbf{Popularity Score} & \textbf{3-gram} & \textbf{Popularity Score} & \textbf{4-gram} & \textbf{Popularity Score} \\ \hline
\verb|[35, 40]| & 1.0638 & [35, 40, 33] & 0.3211 & [40, 33, 34, 44] & 0.1096 \\
\verb|[40, 33]| & 0.5462 & [40, 33, 34] & 0.1861 & [35, 40, 33, 34] & 0.0987 \\
\verb|[18, 5]| & 0.3350 & [35, 40, 0] & 0.1840 & [35, 40, 33, 16] & 0.0656 \\
\verb|[33, 34]| & 0.3129 & [33, 34, 44] & 0.1625 & [35, 40, 0, 42] & 0.0477 \\
\verb|[34, 44]| & 0.2926 & [40, 33, 16] & 0.1083 & [40, 33, 16, 34] & 0.0355 \\
\verb|[5, 40]| & 0.2917 & [0, 42, 20] & 0.0990 & [35, 0, 42, 20] & 0.0320 \\
\verb|[20, 23]| & 0.2764 & [8, 35, 40] & 0.0971 & [35, 40, 0, 33] & 0.0298 \\
\verb|[2, 5]| & 0.2424 & [35, 40, 16] & 0.0757 & [35, 40, 16, 11] & 0.0252 \\
\verb|[23, 18]| & 0.2276 & [18, 35, 40] & 0.0632 & [8, 35, 40, 33] & 0.0240 \\
\verb|[35, 0]| & 0.2261 & [5, 40, 33] & 0.0628 & [0, 42, 20, 23] & 0.0239 \\
\verb|[40, 0]| & 0.2257 & [40, 0, 42] & 0.0602 & [40, 0, 42, 20] & 0.0238 \\
\verb|[6, 5]| & 0.2231 & [33, 16, 34] & 0.0596 & [35, 40, 33, 41] & 0.0222 \\
\verb|[22, 5]| & 0.2208 & [35, 0, 42] & 0.0557 & [40, 33, 16, 11] & 0.0209 \\
\verb|[5, 14]| & 0.2203 & [35, 5, 40] & 0.0541 & [33, 16, 34, 44] & 0.0174 \\
\verb|[29, 18]| & 0.2144 & [35, 40, 24] & 0.0536 & [33, 34, 44, 4] & 0.0168 \\
\verb|[36, 29]| & 0.2127 & [40, 33, 41] & 0.0525 & [5, 40, 33, 34] & 0.0163 \\
\verb|[8, 35]| & 0.2103 & [40, 16, 11] & 0.0524 & [18, 35, 40, 33] & 0.0155 \\
\verb|[5, 18]| & 0.2101 & [2, 35, 40] & 0.0504 & [15, 35, 40, 33] & 0.0150 \\
\verb|[15, 5]| & 0.2048 & [27, 9, 20] & 0.0496 & [35, 40, 33, 42] & 0.0148 \\
\verb|[5, 22]| & 0.2044 & [14, 35, 40] & 0.0466 & [35, 5, 40, 33] & 0.0135 \\
\verb|[16, 11]| & 0.2030 & [42, 20, 23] & 0.0457 & [41, 27, 9, 20] & 0.0114 \\
\verb|[5, 26]| & 0.1953 & [36, 29, 18] & 0.0438 & [0, 33, 34, 44] & 0.0110 \\
\verb|[9, 20]| & 0.1932 & [35, 40, 42] & 0.0437 & [14, 35, 40, 33] & 0.0107 \\
\verb|[27, 9]| & 0.1931 & [41, 27, 9] & 0.0424 & [33, 34, 44, 23] & 0.0106 \\
\verb|[0, 42]| & 0.1931 & [6, 35, 40] & 0.0393 & [40, 0, 33, 34] & 0.0094 \\
\verb|[13, 5]| & 0.1920 & [15, 35, 40] & 0.0391 & [5, 40, 33, 16] & 0.0091 \\
\verb|[5, 46]| & 0.1747 & [26, 35, 40] & 0.0389 & [35, 40, 33, 27] & 0.0091 \\
\verb|[26, 5]| & 0.1731 & [35, 40, 1] & 0.0384 & [6, 35, 40, 33] & 0.0091 \\
\verb|[42, 20]| & 0.1723 & [9, 32, 20] & 0.0376 & [8, 35, 40, 0] & 0.0089 \\
\verb|[5, 36]| & 0.1658 & [22, 35, 40] & 0.0367 & [35, 40, 16, 34] & 0.0088 \\
\hline
\end{tabular}
\caption{Selected words in vocabulary with popularity score.}
\label{table:your_table_label}
\end{table*}

\end{document}